\documentstyle[preprint,pre,aps,eqsecnum]{revtex}

\begin{document}
\draft

\title{Calculations for populations of selected isotopes in
intermediate energy heavy ion collisions}

\author{A. Majumder, and S. Das Gupta}

\address{
Physics Department, McGill University, 3600 University St., Montr{\'e}al\\
Canada H3A 2T8}

\date{ \today}

\maketitle

\begin{abstract}
We compute the populations of isotopes of Boron, Carbon and Nitrogen
measured experimentally in intermediate energy heavy ion collisions.
A two component soluble statistical model is used to find the
initial populations of different nuclei at a finite temperature.
These initial populations are both in particle stable and particle
unstable states.  The particle unstable states then decay.  The
final populations after these decays are computed and compared
with experimental data.
\end{abstract}


\section{INTRODUCTION}

In this work we attempt to calculate the populations of various
isotopes of Boron, Carbon and Nitrogen that were measured in a
number of experiments at the NSCL-MSU facility[1-2].  The calculation
proceeds in two stages.  In the first, primary populations are calculated
in a two-component statistical model.  The calculations in the
first part are exact although numerical.  These populations are
both in particle stable and unstable states.  In the second
stage the particle unstable states are allowed to decay.  This
is done in a Weisskopf formalism.  Exact calculations are very long
and some approximations had to be introduced.  These approximations
will be discussed.  After the decays, the populations are compared
with experiments.

One motivation for this calculation was that it serves as an 
application of the two-component statistical model where an
exact calculation can be done.  This, therefore, could serve as
a benchmark of how far one can trust the predictions of the model.
Unfortunately, the particular predictions we are looking for are also
affected by the subsequent decays.  This effect is not small.
Hence, the predictions are the result of the combination of
two models which had to be applied in tandem before experimental
data could be compared.  A recent application of the 
two-component model was the computation of the caloric curve 
\cite{bha99} in nuclei.

The sections are organised as follows.  Section II gives a brief
description of the two-component statistical model.  After 
presenting, in section III, in words and simple formulae, the overview of the 
secondary decay calculation, we present in section IV the
formalism that we use to model secondary decay.  In section V
we present some calculational details, section VI presents the results
of the calculation.  Summary  and conclusion are presented in
section VII. A short appendix of the more complicated formulae are presented in
section VIII.

\section{THE TWO COMPONENT SOLUBLE STATISTICAL MODEL.}

For completeness, we present here the essential details of 
the two-component statistical model.  The one component model 
was described elsewhere
\cite{das98,das99}.  The
formalism of the two-component model can also be found in 
\cite{bha99}

Assume that the system which breaks up after two heavy ions hit
each other can be desribed as a hot, equilibrated nuclear system
characterised by a temperature $T$ and a freeze-out volume $V$
within which there are $A$ nucleons ($A=Z+N$).  The partition
function of the system is given by

\begin{eqnarray}
Q_{Z,N}=\sum \Pi_{i,j} \frac{ \omega_{i,j} ^{n_{ \small{i,j} }} }{ n_{i,j}! }
\end{eqnarray}

Here $n_{i,j}$ is the number of composites with proton number
$i$ and neutron number $j$, and $\omega_{i,j}$ is the partition
function of a single composite with proton, neutron numbers 
$i, j$ respectively.  There are two constraints: $\sum_{i,j}
in_{i,j}=Z$ and $\sum_{i,j}jn_{i,j}=N$.  These constraints would
appear to make the computation of $Q_{Z,N}$ prohibitively 
difficult, but a recursion relation exists \cite{bha99,das98}
 which allows numerical computation
of $Q_{Z,N}$ quite easy even for large $Z$ and $N$.  Three
equivalent recursion relations exist, any one of which could be
used.  For example, one such relation is

\begin{eqnarray}
Q{z,n}=\frac{1}{z}\sum_{i,j}i\omega_{i,j}Q_{z-i,n-j}
\end{eqnarray}

All nuclear properties are contained in $\omega_{i,j}$.  It
is given by

\begin{eqnarray}
\omega_{i,j}=\frac{V_f}{\hbar^3}(\frac{mT}{2\pi})^{3/2}(i+j)^{3/2}\times
q_{i,j,int}
\end{eqnarray}

Here $V_f$ is the free volume within which the particles move; $V_f$
is related to $V$ through $V_f=V-V_{ex}$ where $V_{ex}$ is the 
excluded volume due to finite sizes of composites.  
We take $V_f$ to be a variable of the
calculation, it is set to be equal to  $fV_0$  where $V_0$ is the
normal volume for $(Z+N)$ nucleons, 
$f$ is then varied to obtain the best fit with
experimental data.  The quantity $q_{i,j,int}$ is 
the internal partition function of the composite.

\begin{eqnarray}
q_{i,j,int}=\sum_{k}^{E_{max}} (2J_k+1)e^{(-E_k/T)}  + q_{i,j,cont} \label{qreal}
\end{eqnarray}

\noindent Where the summation on the right hand side is the contribution from the
discrete spectrum(The cut-off $E_{max}$ is simply the highest energy level that has been
resolved for the given nucleus and is available from data tables); and
$q_{i,j,cont}$ is the contribution from the continuum.
Without loss of generality we can write

\begin{eqnarray}
q_{A,int}=\int \rho_A(E)e^{-\beta E}dE
\end{eqnarray}

\noindent where we have used the abbreviation $A=i+j$, to stand for both $i$ and $j$;
 $\rho_A(E)$ is usually partly discrete and partly continuous.

We will need both $q_{A,int}$ and $\rho_A(E)$.  Volumes of work are
available on $\rho_A(E)$.  This is dealt with in detail in appendix 2B
of \cite{bohr}. The saddle-point approximation for the density of states
assuming a Fermi-gas model is (see eq. 2B-14 in \cite{bohr}) 

\begin{eqnarray}
\rho_{A}(E)=\rho^{0}_{A}(E)\times \exp(ln z_{gr}-\alpha_{0}A+\beta_{0}E)
\end{eqnarray}

For explanations of how $\alpha_0$ and $\beta_0$ are to be chosen see
\cite{bohr} .  In the Fermi-gas model the quantity which is exponentiated is
simply the total entropy $S=As$.  Thus the density of states is given by
a familiar expression $\rho_A(E)=\rho^{0}_{A}(E)\exp(S)$ where $\rho^{0}_{A}(E)$ is the 
pre-factor.  Approximate values of
$\rho^{0}_{A}(E)$ are known provided one does not have to concern with very low value
of $E$ (which we do need).  At temperatures we will be concerned
with, $\exp(S)$ in the Fermi-gas model is given quite accurately by 
$\exp[\pi(\frac{AE}{\epsilon_F})^{1/2}]$.

In the bulk of this paper we adopt this prescription.  For upto $^{20}F$
we write the density of state as $\rho_{A}(E)=\rho^{0}_{A}\times \exp(S)$, where the
low temperature Fermi-gas expression for $S$ as written above is used.  The 
energy independent value of the pre-factor is fixed from experimentally
known levels: 

\begin{eqnarray}
\sum_{k=0}^{E_{max}}(2J_K+1)e^{-E_k/T}=\rho^{0}_{A} \int_0^{E_{max}}e^{(S(E)-\beta E)}dE
\end{eqnarray}

While objections can be raised against this procedure, it achieves
three objectives which we wanted to have: (a) we did not want to lose all
information of the experimentally measured discrete 
excited states ; (b) we did want to take into account the contribution from 
the continuum and (c) with this procedure calculations are fairly simple.
Although, we will not report on all other formulae for density of states
that we also used, our final results for the isotope populations are
quite stable within reasonable variations that were tried.

 We estimate the continuum
contribution as a similar integral from $E_{max}$ to infinity. 

\begin{equation} 
q_{i,j,cont} = \int_{E_{max}}^{\infty} \rho^{0}_{A}\rho_{A} (E) e^{-\beta E }dE
\end{equation}

This process is continued upto $^{20}F$ wherein we can read off energy 
levels from data tables. For elements above $^{20}F$, a 
parametrised version was used, which is given as  

\begin{equation}
q_{i,j,int} = \exp {\left[ \left( W_{0}(i+j) - \sigma (i+j)^{2/3} - \kappa
\frac{i^{2}}{ (i+j)^{1/3} } - s \frac{(j-i)^{2}}{j+i} + T^{2}(i+j)/e \right)/T \right] } \label{qint}
\end{equation}

\noindent where $W_{0}=15.8MeV$, $\sigma = 18.0MeV$, $\kappa = 0.72MeV$, 
$s=23.5MeV$ 
and $e = 16.0MeV$. The first four terms in the right hand side of equation(\ref{qint})
arise from a parametrised version of the binding energy of the ground state. The last
term arises from an approximation to the Fermi-Gas formula for level density. This
was also used in \cite{bon95}. For protons and neutrons $q$ is 1.

 The average number of particles of a composite is given by

\begin{eqnarray}
\langle n_{i,j} \rangle =\omega_{i,j}\frac{Q_{Z-i,N-j}}{Q_{Z,N}} \label{primary}
\end{eqnarray}

However, this population is partly over particle stable states and
partly over particle unstable states which will decay into other
nuclei before reaching the detectors.

\vspace{.5cm}

\section{SECONDARY  DECAY.}

In keeping with the way experimental data are presented, we will
compute ratios of yields of different isotopes of Boron, Carbon,
and Nitrogen.  To lowest order one can consider the $\langle n_{i,j} \rangle $
obtained from equation  (\ref {primary}) above, remove the particle unstable fractions,
 and compare them directly
with experiment.  This is shown in the figures as the dotted line with a 
filled triangle plotting symbol. These populations contain only
 particle stable states.

Next we consider decay of the particle unstable states.
We restrict the secondary decay to be due to emission of six species:
neutron, proton, deuteron, $^3$He, triton, and alpha particles.
Any given nucleus $(i,j)$ from a particle unstable state can in
principle go to at most six other nuclei.  As the populations are canonically
distributed among the various energy levels, we can calculate the fraction that are in
particle stable or unstable states. If the fraction of nuclei $(i,j)$ at the first 
stage in unstable states
is $f^{0} _{i,j}$, then the number of nuclei $(i,j)$ left in particle stable states 
at the stage we call `upto single decay' is given by

\begin{equation}
\langle n_{i,j} \rangle ^{1} = (1 - f^{0} _{i,j}) \langle n_{i,j} \rangle + \sum_{a,b}
(1 - f^{1} _{i,j})\frac { \Gamma _{a,b} }{ \Gamma _{T} } f^{0} _{i+a,j+b}\langle 
n_{i+a,j+b} \rangle
\end{equation}

\noindent where $f^{1}_{i,j}$ is the fraction of the once decayed nuclei in unstable
states. We will indicate how to calculate $f^{1}_{i,j}$ in the next section.
$\Gamma _{a,b}$ is the width for emission by $(a,b)$ from $(i+a,j+b)$ and 
$\Gamma_{T}$ is the total width.

We can then take these revised populations $\langle n_{i,j} \rangle^1 $ and 
again compute the ratios.  We label these `upto single decay'. These are reported 
in the plots
as the small dashed line with the diamond plotting symbol. Note: this is just the
 stable fraction of the population after one stage of decay, the actual
  population is possibly  greater. 
  
After the first decay there may be some  fraction in particle unstable states.  These 
can decay, thereby, changing the population of $(i,j)$ to $\langle n_{i,j} 
\rangle^2 $.  If 
we take the ratios now we get what we call `upto double decay', this is denoted by the
dot$-$dashed line and the square plotting symbol. Again at this stage the 
$\langle n_{i,j} \rangle^2 $ represent only the sum of the stable fractions of the 
populations obtained from the initial distribution, single and double decays.   

It is clear the procedure can be continued.  The fraction remaining
in particle unstable states will continue to decrease.  We found
no significant difference between the `upto triple decay' and the `upto quadruple decay'
calculation.  Thus we do not continue beyond. Once again it should be noted that all the
plotted populations, $\langle n_{i,j} \rangle $,  $\langle n_{i,j} \rangle^1 $, 
$\langle n_{i,j} \rangle^2 $, $\langle n_{i,j} \rangle^3 $
etc., quote only the stable fractions at freezeout, after single, double,
 and triple decay respectively.
 
The formalism for the decay calculation is given in the next
section, there the quantities $f _{i,j}$, $\Gamma _{a,b}$ will be calculated in
 somewhat greater detail.  
The reader who is only interested in the final results
could jump to sections VI and VII.

\vspace{.5cm}

\section{THE DECAY FORMALISM. }

As the heated clusters stream out from the hot source, many of them will be in
particle unstable states, these will decay by particle emission, for example, by
emitting a neutron, proton, $\alpha$ particle etc. They will then leave a residue
nucleus which  may  be particle stable or unstable; if it is unstable then it will
decay further into another isotope and this process will continue till the residue 
is produced in a particle stable state.   

The primary calculation assumes that thermal equilibrium is achieved at freezeout; 
if this is true then the number of composites with $i$ protons and $j$ neutrons with an
  energy  in the interval
 $E$ and $E+dE$ is given by the  canonical factor    

\begin{equation} 
dn_{A}(E) = C_{i,j} \rho_{A} (E) e^{-\beta E } dE  \label{canon}
\end{equation}

\noindent Where we have abbreviated $A$ to mean $(i,j)$, and $\rho_{A} (E)$, from section
II., is given as $\rho_{A} (E) = \rho_{A}^{0} \exp(S)$. The multiplicative constant
$\rho_{A}^{0}$ will, henceforth, be absorbed into the overall normalization constant
$C_{i,j}$. Thus from now on the density function is given simply as 

\begin{equation} 
\rho_{A}(E) = \exp \left[ \pi \bigg(\frac{(i+j)E}{\epsilon_{F}} \bigg)^{1/2} \right]
 \label{dens}
\end{equation}

\noindent $C_{i,j}$ is a normalization constant such that

\begin{equation} 
\int_{0}^{\infty}  C_{i,j} \rho_{A} (E) e^{-\beta E } dE  = \langle n_{(i,j)} \rangle
\end{equation}
 
Now of the various levels in a particular nucleus, some will be at a very low
energy and as a result will be stable to any form of particle decay. Those that lie
above an energy 

\begin{equation}
 E_{x,y} = (M_{x,y} + M_{i-x,j-y} - M_{i,j}) + V_{x,y}  \label{exy}
\end{equation}

\noindent will in general be unstable to decay via emission of a particle 
$(x,y)$
(i.e., a particle with neutron number  $y$ and proton number $x$)  ,
where $M_{x,y}$ is the mass of the particle, $M_{i,j}$ is the mass of the
decaying nucleus $(i,j)$, $M_{i-x,j-y}$ is the mass of the residue left
over after decay and $V_{x,y}$ is the coulomb barrier for that particle.             
Note that $(x,y)$ could represent a variety of particles; in
this note we will consider `six' such particles, as mentioned in the 
introduction.

As is evident from equation (\ref{exy}), different particle decays have different 
energy
thresholds. Consider an isotope $(i,j)$, as an example let us take $^{12}C$ $(
i=6,j=6)$. As we start from the ground state level and
move upwards, we will encounter different thresholds. The lowest will be
 the  $^{4}He$ decay   
threshold at an energy $L_{1}=E_{2,2}$ ( in $^{12}C$ it is at $9.6MeV$ 
approximately) , 
 the next higher threshold is for proton decay at $L_{2}=E_{1,0}$ (in 
$^{12}C$ it is at $18.14MeV$ approximately), and so on; we will get different 
thresholds one after the other( note: the order of different thresholds is 
different for different isotopes ). 

 All nuclei of type $(i,j)$ which are formed between the ground state and the 
lowest threshold $L_{1}$, will remain as isotopes $(i,j)$, this number 
is given by 

\begin{equation}
n_{i,j}(0 \leftrightarrow L_{1}) = \int_{0}^{L_{1}} C_{i,j} \rho (E) e^{-\beta
E } dE  \label{stbpop}
\end{equation}

Those that are
formed between $L_{1}$ and the next threshold $L_{2}$, will all 
completely decay by $^{4}He$ emission, and these nuclei will then appear as 
nuclei of type $(i-2,j-2)$ and must be added on to the population of isotope
$(i-2,j-2)$.

Then, those nuclei of type $(i,j)$ which are formed between $L_{2}$
and the next threshold $L_{3}$,  will decay both   
by $^{4}He$ emission and by  proton emission. In the next zone there will 
be three kinds of decay, and so on. We now ask, how many of the initial nuclei 
formed in a particular zone will decay by each of the channels that are available, 
and how many of the residues formed will be stable or unstable?
 
To answer the above questions: we start by writing down the number of 
particles of type $(x,y)$  with
energy between $(\varepsilon,\varepsilon+d\varepsilon)$ that are emitted, in a time
interval between $t$ and $t+dt$, by nuclei 
of type $(i,j)$, lying between an energy $(E,E+dE)$, leaving behind a 
residue nucleus $(i-x,j-y)$ ( we may
alternatively refer to $(x,y)$, $(i,j)$, and $(i-x,j-y)$ by simply their mass numbers
$a$, $A$ and $B$ where $a=x+y$, $A=i+j$ and $B=i-x+j-y$ )
 
\begin{equation} 
d^{3}N_{a} = W(E,\varepsilon)d\varepsilon dt dN(E,t) \label{num1}
\end{equation}   

\noindent where $dN(E,t)$ is the number of nuclei of type $(i,j)$ 
 initially formed at an 
energy $(E,E+dE)$ which are still left undecayed after a time $t$, given by  
 
\begin{equation}
dN(E,t) = C_{i,j} e^{-\Gamma_{T}(E) t} \rho_{A} (E) e^{-\beta E } dE \label{decay1}
\end{equation}

\noindent and  $W(E,\varepsilon)d\varepsilon$ is the
Weisskopf decay probability per unit time \cite{wei37} given by the expression 

\begin{equation}  
W(E,\varepsilon)d\varepsilon = g_{a} \gamma _{a} \varepsilon \sigma_{[a +
B \rightarrow A ] } \frac { \rho_{B}(E - B_{a} - \varepsilon) \rho_{B}^{0}}{ 
\rho_{A}(E) \rho_{A}^{0}} \label{wisko}
\end{equation}

\noindent In equation (\ref{decay1}), $\Gamma_{T}(E) $ is the the total decay 
probability per unit time from  an energy
level $E$ of the isotope $A$. In equation (\ref{wisko}), $g_{a}$ is the spin 
degeneracy factor
of the emitted particle, $\gamma _{a}$ is a constant of a particular decay
\cite{fri83,wei37}, given
 by 

\begin{equation} 
\gamma _{a} = \frac { m_{p} }{ \pi^{2} \hbar^{3} } \times \frac { a(A-a) 
}{A}
\end{equation}

\noindent where $m_{p}$ is the mass of a nucleon. In equation (\ref{wisko}),
$B_{a}=M_{i-x,j-y}+M_{x,y}-M_{i,j}$, is the separation energy of the decay;
$\sigma_{[a + B \rightarrow A ] }$ is the crossection for the reverse 
reaction to occur, ( i.e. $ a + B \rightarrow A $).  
It is given semiclassically  for uncharged particles as

\begin{equation} 
\sigma_{[a + B \rightarrow A ] } = \pi R_{a}^2
\end{equation}

\noindent and for charged particles as 

\begin{equation} 
\sigma_{[a + B \rightarrow A ] } = \pi R_{a}^2 \frac { \varepsilon - V }{ \varepsilon } 
\theta ( \varepsilon -V) \label{cros}
\end{equation}

\noindent where $R_{a}$ is the radius associated with the geometrical crossection  of 
the         
formation of A from B and $a$. Following the prescription of Friedmann and Lynch
\cite{fri83}, $R_{a}$ is given by

\[ 
R_{a} = \left\{ \begin{array}{ll}
 [(A-a)^{1/3} + (a)^{1/3}]r_{0} & \mbox{, for $ a \geq 2 $ } \\ 
 r_{0}(A-1)^{1/3} & \mbox{, for $a=1$ }
\end{array} 
\right. 
\]

\noindent where $r_{0} = 1.2 fm$.
In equation (\ref{cros}), $V$ is the Coulomb barrier for the formation of A from B 
and $a$. Again following \cite{fri83}, this is written in the touching sphere
 approximation as

\[ 
V_{a} = \left\{ \begin{array}{ll}
 \frac {x(i-x)e^{2}}{((A-a)^{1/3} + (a)^{1/3})r_{c}} & \mbox{, for $ a \geq 
 2 $,} \\ 
 \frac {(i-1)e^{2}}{ r_{c}(A)^{1/3}} & \mbox{, for protons, }
\end{array} 
\right. 
\]

\noindent where $r_{c} = 1.44 fm$.
Also in equation (\ref{wisko}), $\rho_{A}(E)$ , $\rho_{B}(E-B_{a}-\varepsilon)$ are 
the respective density of states of the two nuclei. They have the same form as in 
equation (\ref{dens}). Also $\rho_{A}^{0}$ and $\rho_{B}^{0}$ are the respective
multiplicative constants for the density of states, as mentioned in section(II).

We note that $d^{3} N_{a}$ in equation (\ref{num1}) is also equal to the number of
nuclei that were initially formed as nuclei of type $A$ at an energy 
between $E$ and $E+dE$, and then decayed into
  nuclei $B$ with an excitation energy of
$E-B_{a}-\varepsilon$. To get the total number of states that decayed 
from a level $E$ by emission of a particle of any allowed energy, we integrate over 
$\varepsilon$
 from its minimum value $V_{a}$ to its maximum value $E-B_{a}$,  and get 

\begin{equation} 
d^{2}N = \Gamma_{a} (E) C_{i,j} e^{-\Gamma_{T} (E) t} \rho_{A} (E) e^{-\beta E } dEdt 
\end{equation}
 
\noindent where

\begin{equation} 
\Gamma_{a} (E) = \int _{V_{a}}^{E-B_{a}} W(E,\varepsilon) d\varepsilon \label{befb}
\end{equation} 

\noindent which on integration gives 

\begin{eqnarray} 
\Gamma_{a} (E) &=& \frac { 2 \gamma_{a}\prime \rho_{B}^{0}}{ \rho_{A}(E) \rho_{A}^{0}}
 \left[ (E-B_{a}-V_{a})
\left(
\frac { e^{C {\cal B} } }{C} ( {\cal B} - 1/C) - \frac { e ^{C {\cal A} } }{C} 
( {\cal A} - 1/C) \right) \right. \nonumber \\
 & & \left. \mbox{} - \frac { e ^{C {\cal B} }}{C} \left( {\cal B} ^{3} - \frac 
 {3 {\cal B} ^{2} }{C} + 
 \frac { 6 {\cal B} }{C^{2}} - \frac {6}{C^{3}} \right)\right. \nonumber \\
 & & \left. \mbox{} + \frac {e ^{C { \cal A} }}{C} \left( { \cal A} ^{3}-\frac {3
 {\cal A} ^{2}}{C}+\frac {6 { \cal A} }{C^{2}}-
\frac {6}{C^{3}} \right) \right]  \label{big}
\end{eqnarray}

\noindent ( the derivation of the above equation is given in the appendix ), where 
$C = \pi (\frac{i+j-x-y}{\epsilon_{F}})^{1/2} $, ${ \cal B} = \sqrt{E-B_{a}-V_{a}}$
, ${ \cal A} = 0$. In the above equation $g_{a}$ (equation(\ref{wisko})), and some of 
the factors of $\sigma$ (equation(\ref{cros})) have been absorbed into 
$\gamma_{a}^{\prime}$ thus 

\begin{equation} 
\gamma_{a} \prime = \gamma_{a} g_{a} \pi R_{a}^{2}
\end{equation}

\noindent  
We may now integrate out the time to get 

\begin{equation} 
dN_{a} = \frac {\Gamma_{a}(E)}{\Gamma_{T} (E) }  C_{i,j} \rho_{A} (E) e^{-\beta E } dE 
\end{equation}

\noindent To get the total number of states that have decayed from nuclei of type 
$A$ by channel $a$ we must integrate over $E$ from $L_{1}$ to $\infty$,

\begin{equation} 
N_{a} = \int_{L_{1}}^{\infty} \frac {\Gamma_{a}(E)}{\Gamma_{T} (E) }  C_{i,j} \rho_{A}
 (E) e^{-\beta E } dE  \label{num4}
\end{equation}

\noindent This integration is quite involved for as we crossover from one zone of decay
$(L_{1},L_{2})$ to another zone $(L_{2},L_{3})$, $\Gamma_{T} (E)$  changes 
discontinously as a new channel of decay becomes accessible to the nuclei. Thus we break
 up the integration into $6$ zones, corresponding to the $6$ real decay zones, and 
 integrate within each zone independently. Note that the last zone extends from $L_{6}$
 to $L_{7}=\infty$, and is thus considerably larger than the other zones. However, at the 
 low temperatures that will be encountered, this zone will be sparsely populated. 
 Thus the following approximation is valid. 
 Within each zone, with an energy from
$L_{k}$ to $L_{k+1}$, the integral can be replaced by a mean value expression,

\begin{equation} 
N_{a}(L_{k},L_{k+1}) = \frac { _{L_{k}} \langle \Gamma_{a} \rangle _{L_{k+1}} }{ 
_{L_{k}}
\langle \Gamma_{T} \rangle _{L_{k+1}} } \Delta n_{i,j} (L_{k},L_{k+1}) \label{num5}
\end{equation}

\noindent Where $N_{a}(L_{k},L_{k+1})$ is the mean number of nuclei of type $(i,j)$ (or
$A$) that were initially formed at an energy between $L_{k}$ and $L_{k+1}$, 
and decayed by the $(x,y)$ (or $a$) channel. In the above equation 
 
\begin{equation} 
_{L_{k}} \langle \Gamma_{a} \rangle _{L_{k+1}} = \int_{L_{k}}^{L_{k+1}} \Gamma _{a} (E)
C_{i,j} \rho_{A} (E) e^{-\beta E } dE  \label{gamma_lk}
\end{equation}

\noindent Of course, the left hand side is zero if channel $a$ is not open in the region
$L_{k}$ to $L_{k+1}$. The mean decay rate over all channels is  

\begin{equation} 
_{L_{k}} \langle \Gamma_{T} \rangle _{L_{k+1}} = \int_{L_{k}}^{L_{k+1}} \Gamma _{T} (E)
C_{i,j} \rho_{A} (E) e^{-\beta E } dE  \label{num6}
\end{equation}

\noindent and 

\begin{equation}   
\Delta n_{i,j} (L_{k},L_{k+1}) = \int_{L_{k}}^{L_{k+1}} C_{i,j} \rho_{A} (E) e^{-\beta
E } dE \label{num7}
\end{equation}

\noindent Thus by summing up all the contributions from the six different zones, 
we get the total number of nuclei that have decayed from isotope $A$ by the 
$a$  channel as
 
\begin{equation}   
N_{a} = \sum_{k=1}^{6} N_{a}(L_{k},L_{k+1})  \label{AtoB}
\end{equation}

 To find out how many of these have decayed to stable isotopes, we must first 
calculate from  equation (\ref{num1}) the stable decay rate $\Gamma_{a}^{s}(E)$.
Two cases emerge in this calculation. If $E-B_{a}-V_{a} \geq E_{A-a}^{s}$, 
$\Gamma_{a}^{s}(E)$ is obtained by integrating over $\varepsilon$,  
from  $(E-B_{a}-E_{A-a}^{s})$ to
  its maximum value $(E-B_{a})$, where  $E_{A-a}^{s}$ is the stable level or the lowest
   threshold 
$L_{1}$ of the residue nucleus $B$ above which $B$ is 
unstable. The expression for 
$\Gamma_{a}^{s}(E)$ is obtained from that of $\Gamma_{a}(E)$ in equation (18) by
replacing ${\cal B} = \sqrt{E_{A-a}^{s}} $ .  If $E-B_{a}-V_{a} < E_{A-a}^{s}$,
then $\Gamma_{a}^{s}(E) = \Gamma_{a}(E)$. Then, following a similar 
procedure as above for $\Gamma_{a}$, we get the total 
number of nuclei $A$ (or $(i,j)$) lying in an energy range between 
$(L_{k},L_{k+1})$, that 
decay by the $a$ channel to a stable state as

\begin{equation}   
N_{a}^{s}(L_{k},L_{k+1}) = \frac { _{L_{k}} \langle \Gamma_{a} ^{s} \rangle _{L_{k+1}} 
}{ _{L_{k}} \langle \Gamma_{T} \rangle _{L_{k+1}} } \Delta n_{i,j} (L_{k},L_{k+1})
\label{sdert}
\end{equation}

The unstable decay rate from a particular level or zone is the probability of a 
decay
per unit time from $A$ to an unstable level or levels of $B$ from which further 
decay can take place. It is
easy to see that they are given simply as the difference of the total decay rate and 
the
stable decay rate i.e., 

\begin{equation}   
\Gamma_{a} ^{u} = \Gamma_{a} - \Gamma_{a}^{s}
\end{equation}

\noindent The derivations and expressions for the full decay rates are given in the    
appendix.

After a decay has taken place ( $ A \rightarrow B + a $),
we ask what is the population distribution of the residue
as a function of its energy ( $x = E- B_a - \varepsilon $ ). This can, 
in principle, be
calculated from equation(\ref{num1}) by integrating over $E$ and 
$\varepsilon$, such
that $(x = E- B_{a} - \varepsilon )$, the energy of the residue, is a constant.
 First we make a
change of variables from $(E ,\varepsilon )$ to $(E,x)$ and then integrate over $E$
only. We get                  

\begin{equation}   
dN_{a}(x) = \left( \int _{B_{a}+V_{a}+x}^{\infty} dE \gamma_{a}^{\prime}  \frac{
E-B_{a}-V_{a}-x}{\Gamma_{T}(E)} \rho_{B}(x) \frac{\rho_{B}^{0}}{\rho_{A}^{0}}
C_{i,j}e^{ - \beta E} \right) dx
\label{befshady}
\end{equation}

\noindent This integration is quite involved. We assume that the residue
population is canonically distributed, but with a new temperature 
$1/\beta^{\prime}$ i.e.,

\begin{equation}   
dN_{a}(x) = {\cal D}_{i,j \rightarrow k,l }  \rho_{B}(x) e^{- \beta^{\prime} x } \label{canres}
\end{equation}

\noindent There are two unknowns in this formula, the new temperature $1/\beta^{\prime}$ 
and
the overall normalization constant ${\cal D}_{i,j \rightarrow k,l}$. To find these two
constants we will impose that the total population of this interim stage (i.e. 
$N_{i,j \rightarrow k,l }$ ), and the mean energy of the distribution 
$\langle x \rangle$, be reproduced by this new temperature. 

We can obtain formal expressions for the total population of the residue $B$ as
contributed by the decay of $A$, as well as its mean energy $\langle x \rangle$, 
from equation(\ref{canres}) as

\begin{eqnarray}   
N_{a}(\beta^{\prime},{\cal D},C) &=& \int_{0}^{\infty} dN_{a}(x) \nonumber \\
     &=& \frac { {\cal D}_{i,j \rightarrow k,l} }{ \beta^{\prime} } \left[ 1 + C\sqrt{
     \frac {\pi}{4\beta^{\prime}} } e^{ C^{2}/4\beta^{\prime} } \left( 1-erf \left(\frac{ C}{2
     \sqrt{\beta^{\prime}}}\right) \right) \right]  \label{frmlN}
\end{eqnarray}
 
\begin{eqnarray}   
\langle x(\beta^{\prime},C) \rangle &=& 
\frac{1}{N_{a}(\beta^{\prime},{\cal D},C)} 
\int_{0}^{\infty} x dN_{a}(x) \nonumber \\
 &=& {\cal D}_{i,j \rightarrow k,l} \Bigg[ \frac{1}{{\beta^{\prime}}^{2}} +
 \frac{3\sqrt{\pi}C}{{\beta^{\prime}}^{5/2}} \Bigg\{ 1 +
 erf\left(\frac{C}{2\sqrt{\beta^{\prime}}}\right)
 \Bigg\}e^{C^{2}/4\beta^{\prime}} \nonumber \\
 & & \mbox + \frac{C^{2}}{4{\beta^{\prime}}^{3}} + \frac{C^{3}\sqrt{\pi}}{
 8{\beta^{\prime}}^{7/2}}\Bigg\{ 1
 + erf\left(\frac{C}{2\sqrt{\beta^{\prime}}}\right)\Bigg\}e^{C^{2}/
 4\beta^{\prime}} \Bigg] \label{frmlx}
\end{eqnarray}

Where the formal expression for $N_{a}(\beta^{\prime},{\cal D},C)$ 
is used in equation(\ref{frmlx}).
 The numerical value of $N_{a}$ is taken from
equation(\ref{AtoB}). The numerical value of $\langle x \rangle$ is 
found by explicit use of equation (\ref{befshady}). From these two 
equations we obtain the two constants ${\cal D}_{i,j \rightarrow k,l } $ 
and $\beta^{\prime}$. 

The numerical value of $\langle x \rangle$ is derived from equation (\ref{befshady})
as follows. 

\begin{equation}   
\langle x \rangle = \frac{1}{N_{a}} \int_{0}^{\infty} dx x 
\left( \int _{B_{a}+V_{a}+x}^{E_{max}} dE \gamma_{a}^{\prime} 
\frac{E-B_{a}-V_{a}-x}{\Gamma_{T}(E)} \rho_{B}(x) \frac{\rho_{B}^{0}}{\rho_{A}^{0}}
 C_{i,j}e^{ - \beta E} \right)
\end{equation}

\noindent In the above equation, the numerical value of $N_{a}$ is taken from 
equation(\ref{AtoB}). We may now change the order of integration to get 

\begin{equation}   
\langle x \rangle = \frac{1}{N_{a}} \int_{B_{a}+V_{a}}^{\infty} dE 
\int_{0}^{E-B_{a}-V_{a}} dx x \gamma_{a}^{\prime} 
\frac{E-B_{a}-V_{a}-x}{\Gamma_{T}(E)} \rho_{B}(x) \frac{\rho_{B}^{0}}{\rho_{A}^{0}}
C_{i,j}e^{ - \beta E}
\end{equation}

\noindent The $x$ integration is now done simply to obtain 

\begin{equation}   
\langle x \rangle = \frac{1}{N_{a}} \int_{B_{a}+V_{a}}^{\infty} dE \gamma_{a}^{\prime}
\frac{I(E)}{\Gamma_{T}(E)} C_{i,j} \rho_{A}(E) e^{ -\beta E}
\end{equation}

\noindent where $I(E)$ is given by 

\begin{eqnarray}
I(E) &=& \frac{1}{\rho_{A}(E)} \frac{\rho_{B}^{0}}{\rho_{A}^{0}}\Bigg[ 
\frac{4(E-B_{a}-V_{a})^{2} e^{C\sqrt{E-B_{a}-V_{a}}} }{ C^{2} } -
\frac{28(E-B_{a}-V_{a})^{3/2} e^{C\sqrt{E-B_{a}-V_{a}}} }{ C^{3} } \nonumber \\
& & \mbox{} + \frac{108(E-B_{a}-V_{a}) e^{C\sqrt{E-B_{a}-V_{a}}} }{ C^{4} } 
  - \frac{240(E-B_{a}-V_{a})^{1/2} e^{C\sqrt{E-B_{a}-V_{a}}} }{ C^{5} } \nonumber \\
& & \mbox{} + \frac{240 e^{C\sqrt{E-B_{a}-V_{a}}} }{ C^{6} } + 
\frac{12(E-B_{a}-V_{a})}{C^{4}} - \frac{240}{C^{6}} \Bigg]   
\end{eqnarray}

In the ensuing integration over $E$, we, once again, replace the integral with 
its mean value expression. 

\begin{equation}   
\langle x \rangle =  \frac{1}{N_{a}}\int_{B_{a}+V_{a}}^{\infty} dE \gamma_{a}^{\prime}
\frac{_{B_{a}+V_{a}}\langle I(E) \rangle_{\infty} }{_{B_{a}+V_{a}}\langle 
\Gamma_{T}(E) \rangle_{\infty} } C_{i,j} \rho_{A}(E) e^{ -\beta E} \label{actlx}
\end{equation}

\noindent where 

\begin{equation}   
_{B_{a}+V_{a}}\langle \Gamma_{T}(E) \rangle_{\infty} = \sum_{E_{k} > B_{a} + V_{a}}
\mbox{}_{L_{k}}\langle \Gamma_{T}(E) \rangle_{L_{k+1}}  
\end{equation}

\noindent and

\begin{eqnarray}   
_{B_{a}+V_{a}}\langle I(E) \rangle_{\infty} &=&  \int_{B_{a}+V_{a}}^{\infty} dE
I(E) C_{i,j} \rho_{A}(E) e^{ -\beta E} \nonumber \\ 
&=& C_{i,j} e^{-\beta G_{a}} \frac{\rho_{B}^{0}}{\rho_{A}^{0}} \Bigg[
\frac{1}{\beta^{4}} + \frac{3C\sqrt{\pi}}{4 \beta^{9/2}} \Bigg\{1 + erf
\left(\frac{C}{2\sqrt{\beta}}\right) \Bigg\}e^{C^{2}/4\beta} \nonumber \\
 & & \mbox + \frac{C^{2}}{4\beta^{5}} + \frac{C^{3}\sqrt{\pi}}{8\beta^{11/2}}\Bigg\{ 1
 + erf\left(\frac{C}{2\sqrt{\beta}}\right)\Bigg\}e^{C^{2}/4\beta} \Bigg]
\end{eqnarray}

Thus the formal expressions for $N_{a}(\beta^{\prime},{\cal D},C)$ 
( equation(\ref{frmlN})), and    
$\langle x(\beta^{\prime},C) \rangle$ (equation(\ref{frmlx})), are compared to the actual values
obtained for $N_{a}$ (equation(\ref{AtoB})), and $\langle x \rangle$
(equation(\ref{actlx})), and the two unknowns of equation(\ref{canres}) are evaluated.
We can now proceed with further decays following the same procedure as before with
decay occurring from a canonically distributed population at a temperature
$1/\beta^{\prime}$.
 
We can thus model an n$-$step decay process by assuming that at each intermediate stage
the population is canonically distributed with a new temperature and overall
normalization constant. The decay rates to the next stage are calculated with the new
temperature. Following this, the fraction of the population that decays through 
a particular channel,
and the mean energy of the resultant residue nucleus, are calculated. These are then
used to secure the temperature and normalization constant of the next stage of decay.
This process will continue till the fraction of decay to particle unstable states
becomes negligible. 

\vspace{1cm}
\section{ THE CALCULATION.  }

From the primary calculation we obtain that $\langle n_{i,j} \rangle$ nuclei of 
type $(i,j)$ ( or $A$) are formed from the
initial multifragmentation. The population $\langle n_{i,j} \rangle$ is distributed
canonically among the various energy levels 
as demonstrated by equation(\ref{canon}). If a particular nucleus is at a sufficiently 
excited state
then it will emit a particle  $(x,y)$ (or $a$) and leave a residue $(i-x=k,j-y=l)$ 
(or $B$),
which may again decay by emitting a particle $(u,v)$ (or $b$) leaving a nucleus
$(k-u=m,l-v=n)$ ( or $D$ ), and 
so on untill it finally reaches a
nucleus $(p,q)$ (or $Z$) in a stable state. We ask the question that if 
$\langle n_{i,j}    
\rangle$ nuclei of type $A$ were intially formed, then how many of these will
finally end up as stable nuclei of type $A$, $B$, $D$ 
 ... $Z$. The  contribution of $\langle n_{i,j} \rangle$ to the final stable 
  population of $A$ is given simply by equation(\ref{stbpop}) as 
 
\begin{equation}
n_{A}^{f}=\int_{0}^{L_{1}} C_{i,j} \rho_{A} (E) e^{- \beta E } dE = \Delta n_{i,j}
(0,L_{1})
\label{stpop}
\end{equation}   
  
\noindent the number of nuclei initially formed as $(i,j)$ which  decay to 
$(i-x,j-y)=(k,l)$ is given as  

\begin{equation}
n_{A \rightarrow B} = \sum _{k=1}^{6} \frac { _{L_{k}} \langle \Gamma_{a} \rangle 
_{L_{k+1
}} }{ _  {L_{k}} \langle \Gamma_{T} \rangle _{L_{k+1}} } \Delta n_{i,j} (L_{k},L_{k+1})
\end{equation} 
 
\noindent The mean energy of the newly formed residue nucleus is given by 

\begin{equation}
\langle x \rangle =  \frac{1}{n_{A \rightarrow B}}\gamma_{a}^{\prime}
\frac{_{B_{a}+V_{a}}\langle I(E) \rangle_{\infty} }{_{B_{a}+V_{a}}\langle 
\Gamma_{T}(E) \rangle_{\infty} } \Delta n_{i,j} (B_{a}+V_{a},\infty) 
\end{equation} 

We assume that this population is canonically distributed from an excitation energy
 of $E_{0}=0$ to $\infty$ with a new temperature $1/\beta^{\prime}$( equation \ref{canres}).
Extraction of the new temperature $1/\beta^{\prime}$ and the overall normalization
constant ${\cal D}_{i,j\rightarrow k,l}$ is done as detailed in section(IV).
In most cases, where this 
procedure was implemented, we obtained a new temperature $1/\beta^{\prime}$ which was 
lower than the
initial temperature $1/\beta$; however, in about $3\%$ of the cases $1/\beta^{\prime}$
turned out to be higher than $1/\beta$; this occurs when the residue of the decay
process is far from the valley of stability.  
We can then calculate the number of nuclei that initially started out as $A$'s
 and finally ended up as `stable' $B$'s as 

\begin{equation}
n_{A \rightarrow B}^{f} = \int_{0}^{L_{1}} {\cal D}_{i,j \rightarrow k,l}
 \rho_{B} (x) e^{-\beta^{\prime} x } dx
\end{equation} 

\noindent note that in the above equation $\rho_{B} (x)$ and $L_1$ are the density of 
states
and lowest decay threshold for the nucleus of type $(i-x=k,j-y=l)$. This number can
 also be calculated directly by using the stable decay rates (equation(\ref{sdert})) 

\begin{equation}
n_{A \rightarrow B}^{f} = \sum _{k=1}^{6} \frac { _{L_{k}} \langle \Gamma_{a} ^{s} 
\rangle _{L_{k+1
}} }{ _  {L_{k}} \langle \Gamma_{T} \rangle _{L_{k+1}} } \Delta n_{i,j} (L_{k},L_{k+1})
\end{equation} 

\noindent The second equation is more correct as it does not depend on the assumption
 that
the residue is canonically distributed. A comparison of the $n_{A \rightarrow B}^{f}$ 
obtained from the above two
equations gives an estimate of the error involved in the assumption of a canonically 
distributed residue population. Now we ask, what is the number of 
nuclei 
of the B's just formed which will decay by
emitting a particle $b$ to a nucleus of type $D$; this is
calculated simply as 

\begin{equation}
n_{A \rightarrow B \rightarrow D} =  \sum _{k=1}^{L_{6}} \frac { _{L_{k}} \langle   
 \Gamma_{b} \rangle 
_{L_{k+1}} }{ _{L_{k}} \langle \Gamma_{T} \rangle _{L_{k+1}} } \Delta 
n_{A \rightarrow B} (L_{k},L_{k+1}) 
\end{equation} 

\noindent the decay rates in the above equation are calculated with the temperature
$1/\beta^{\prime}$. We then calculate the mean energy $\langle y \rangle$ of the 
new distribution
as

\begin{equation}
\langle y \rangle =  \frac{1}{n_{A \rightarrow B \rightarrow D}}\gamma_{b}^{\prime}
\frac{_{B_{b}+V_{b}}\langle I(x) \rangle_{\infty} }{_{B_{b}+V_{b}}\langle 
\Gamma_{T}(x) \rangle_{\infty} } \Delta n_{A \rightarrow B} (B_{b}+V_{b},\infty) 
\end{equation} 

Using these, we continue the process on, by again calculating the temperature and 
norm of a canonical distribution,
which when summed from excitation energy $0$ to $\infty$ is
equal to $n_{A \rightarrow B \rightarrow D}$, and whose mean energy is equal to 
$\langle y \rangle $. 
We can then proceed to find how many of
these will be in stable states, how many will decay on further etc. We continue this
process till the contribution from this decay chain, $A \rightarrow B \rightarrow D       
\rightarrow ... $, will give numbers of nuclei negligible compared to the already 
present number in stable states.

\vspace{1cm}
\section{ RESULTS OF THE CALCULATION.  }

Our objective is to calculate the yields of the Boron,Carbon and Nitrogen isotopes 
 measured in the $S + Ag$ Heavy-Ion collision at an energy of
$22.3AMeV$ \cite{xu89}. In figures 1 to 5 the data are shown as empty squares.
 The method of calculation is simple, first we calculate the
  primary populations of the isotopes using equation(\ref{primary}). We then 
  remove the unstable fraction of the population, and quote only the stable part.
 This is denoted by the dotted line and triangle plotting symbol.
  We then incorporate secondary decay
 by adding on all the populations  of nuclei that can reach
 a stable level of the isotopes by emitting only one of the six particles considered. We
 call these the `upto single decay' populations and denote them by the small dashed
 line and diamond plotting symbol. We then add on all those unstable nuclei which can 
 reach a
stable level of the given isotopes by sequentially emitting any two particles of the six
considered. We call these the `upto double decay' populations and denote them by the 
dot$-$dashed line and square plotting symbol. We then add on all those that can reach
the isotopes by three particle emissions, called the `upto triple decay' population 
and denoted by the large
dashed line and star plotting symbol. And finally we add on the `upto quadruple decay'
population denoted by the solid line and circle plotting symbol. As there is negligible
difference between  `upto triple decay' and `upto quadruple decay' we stop after
`quadruple decay'.

To fit with experimental data, we have four parameters to tune, the obvious ones being
the initial temperature $\beta$ or $T$, the free volume $V_{f}$ of the primary 
calculation, the ratio $A/Z$ 
( as one does not know how much loss due to pre$-$equilibrium emission has taken place )
and an overall multiplicative constant ${\cal H}$ ( as we do not know how many nuclei
collided in the experiment ). The plots are noted to be most sensitive to $\beta$ and 
$A/Z$. Thus in fitting the data we first set particular values of $\beta$ and $A/Z$, 
and
calculate the multiplicities at all stages of decay ( $V_{f}$ is varied to get the best
possible fit at this temperature and $A/Z$ ) . We then multiply all the 
multiplicities by an appropriate ${\cal H}$  and take the logarithm. These are
then plotted and compared with $\log(counts)$ obtained from the experiment. We then vary
$\beta$, $A/Z$ and repeat the above procedure till a good fit is obtained. 
We present fits for
three different temperatures, and different $A/Z$ for each temperature.  
$V_{f}$ and ${\cal H}$ are set to
obtain the best fit possible for a given $\beta$ and  $A/Z$.

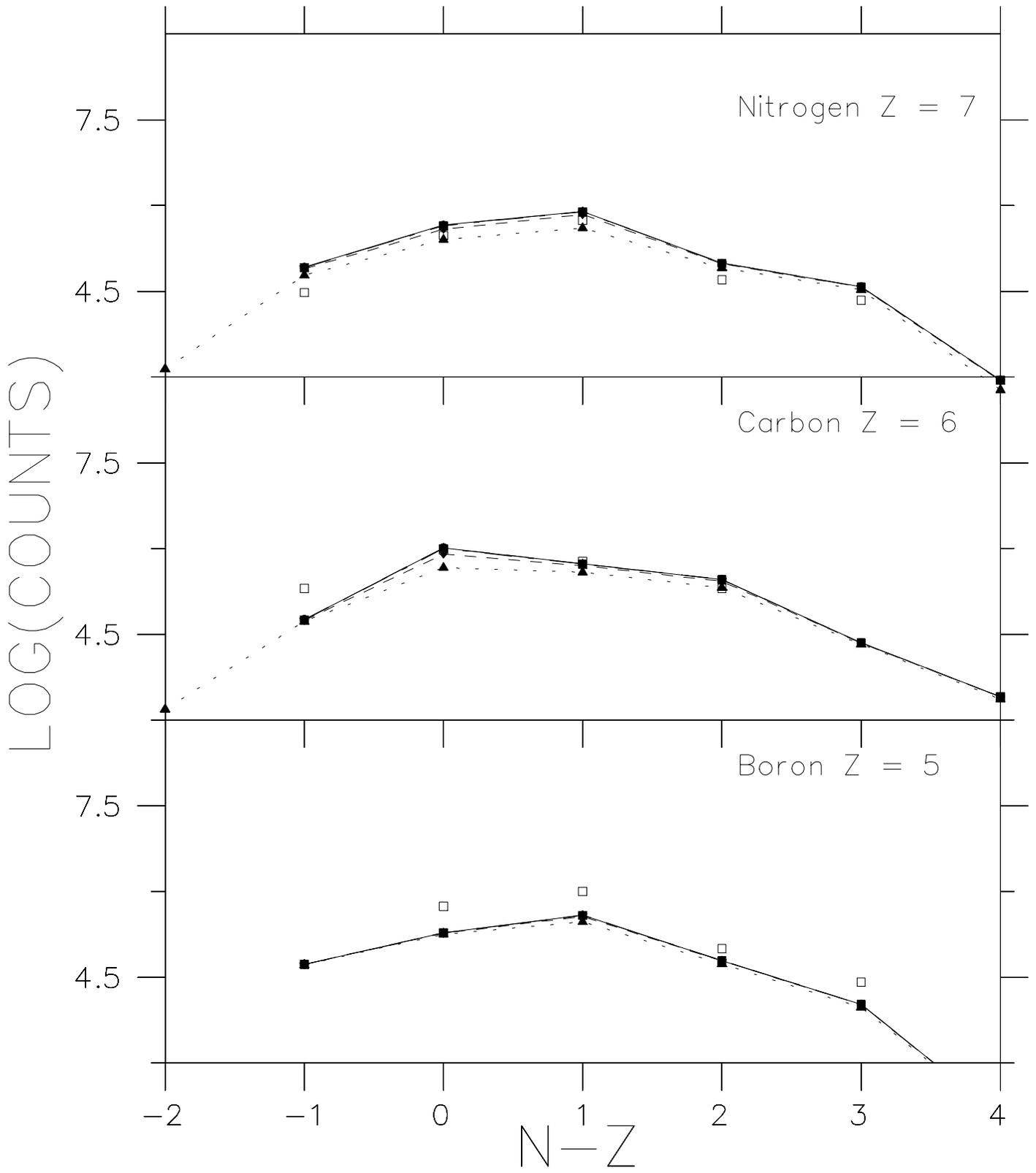
\begin{figure} \caption{ Log(counts) vs Neutron number (N) $-$ Proton Number (Z) for the 
three cases of 
Boron,Carbon and Nitrogen. The experimental data are from [1] S+Ag at $22.3AMeV$.
The fits show varying stages of decay for a total $Z_{T}=50$, $A_{T}=110$, $T=3.0MeV$,
$V_{fr}/V_{0}= 3.0$ and $\log {\cal H } = 6.82$. The empty squares 
 are the experimental data. The dotted line with the triangle plotting symbol is the 
 primary calculation. The small dashed line with diamond plotting symbol is the 
`upto single decay' 
calculation. The dot-dashed line with square plotting symbol is the `upto double decay'
calculation. The  dashed line with star plotting symbol is the `upto triple decay'
calculation. The solid line with circle plotting symbol is the `upto quadruple decay'
calculation.   } \label{fig1} \end{figure}

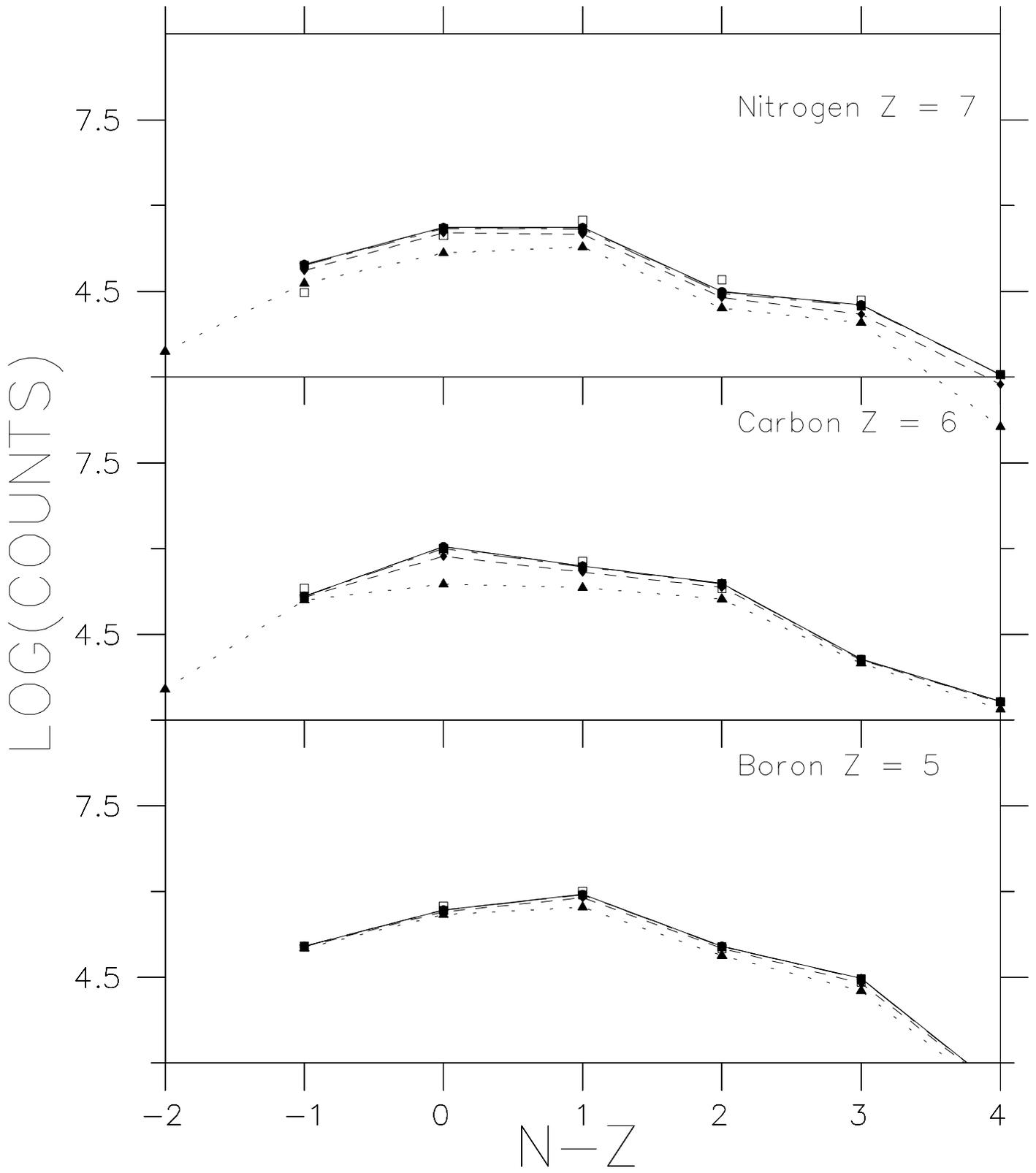
\begin{figure} \caption{ Same as fig.1 but with $Z_{T}=50$, $A_{T}=110$, $T=5.0MeV$, 
$V_{fr}/V_{0}= 5.5$ and $log{\cal H } = 6.34 $. The best fit with the data has been
obtained with these parameters. } \label{fig2} \end{figure}

 We note that the $S+Ag$ system is one with $A=139$ and $Z=63$ thus $A/Z=2.2$. The
 authors of \cite{xu89} state that some pre-equilibrium emission may have taken place.
 As we do not know what proportion of neutrons and protons are lost in such a process, we
 start the calculation with the same $A/Z$ as the $S+Ag$ system.
 We start with a $Z=50$ and $A/Z=2.2$ i.e. $A=110$. We start the calculation with a
 low temperature of $3MeV$ in Figure(\ref{fig1}) ( $V_{f}$ and ${\cal H}$ are varied to
 get the best fit ). We note that overall there is a slight excess of the heavier
 Nirogen isotopes as compared to data and a deficit of the lighter Boron isotopes, this
 implies that the temperature is too low and enough of the light isotopes are not being
 formed. We proceed by raising the temperature to $5MeV$, maintaining the same $A/Z$. By
 now varying $V_{f}$ and ${\cal H}$ we find an excellent fit with the data
 (figure(\ref{fig2})). 
 
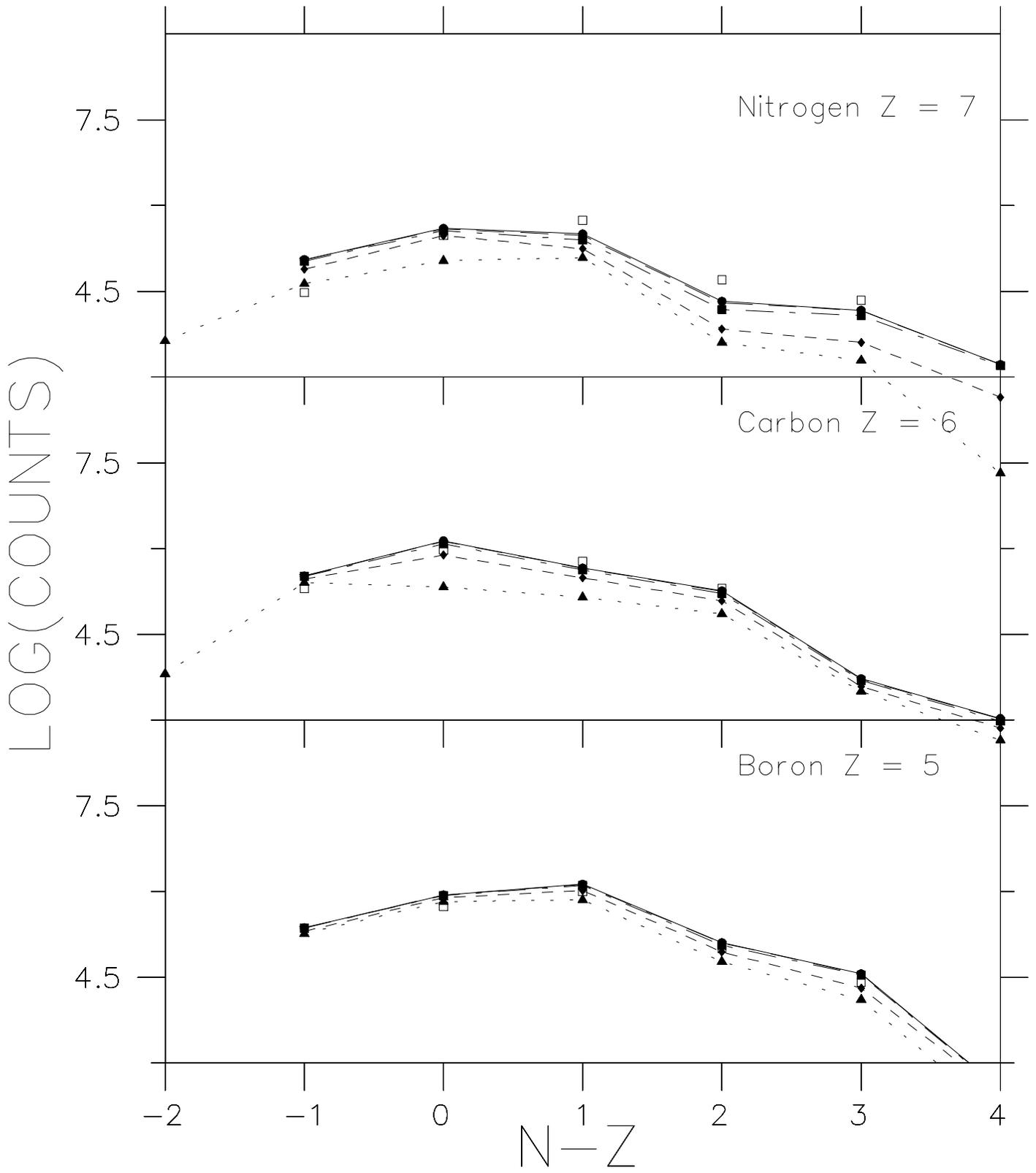
\begin{figure} \caption{ Same as fig.1 but with $Z_{T}=50$, $A_{T}=110$, $T=7.0MeV$,
 $V_{fr}/V_{0}= 3.0$ and $log{\cal H } = 6.82 $. } \label{fig3} \end{figure}

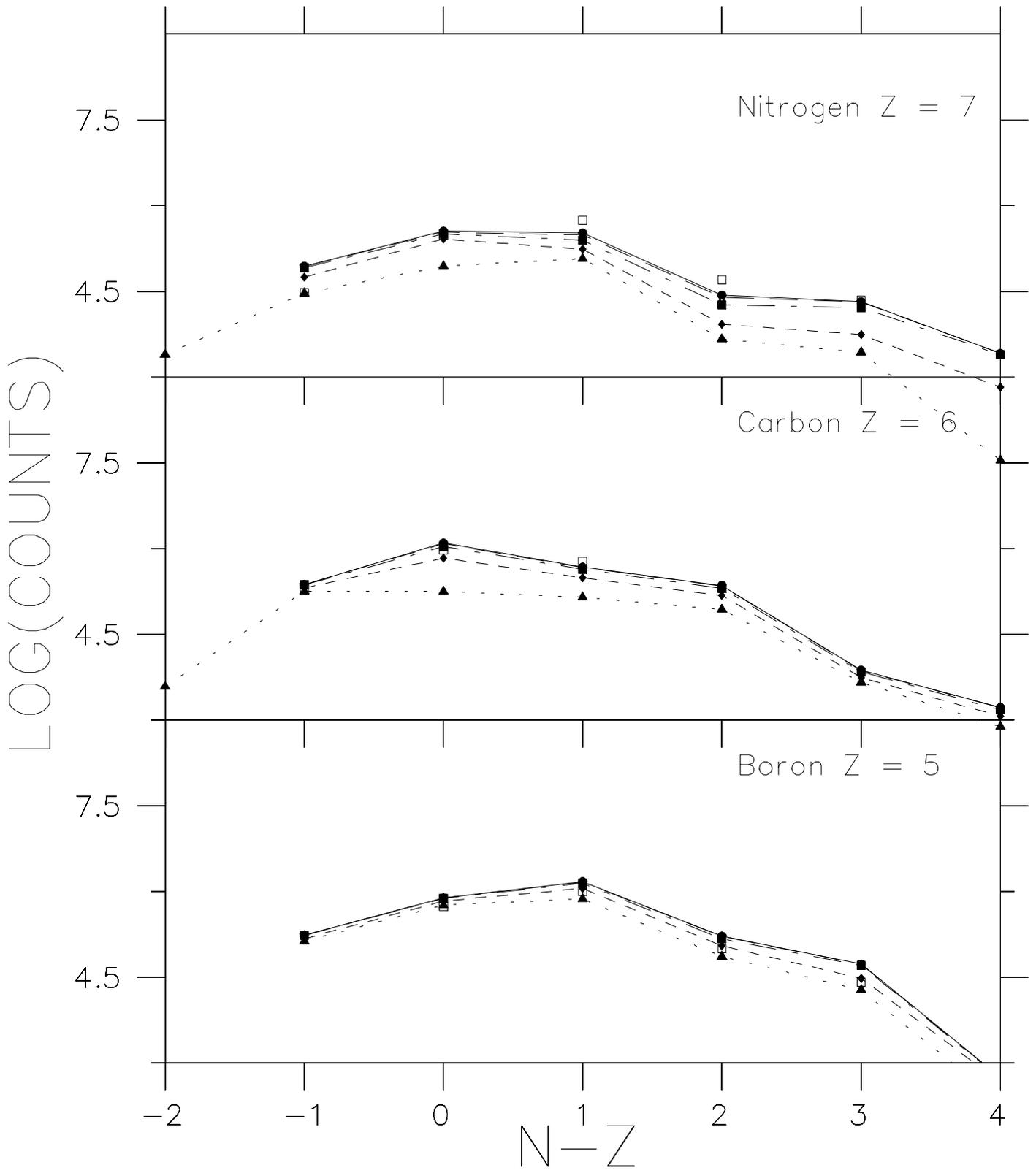
\begin{figure} \caption{ Same as fig.1 but with $Z_{T}=50$, $A_{T}=115$, $T=7.0MeV$, 
$V_{fr}/V_{0}= 3.0$ and $log{\cal H } = 6.82 $. } \label{fig4} \end{figure}

 One may ask at this point, if there is more than one set of parameters which fits the data
 well. To answer this question we increase, first, the temperature to $7MeV$, maintain the
 same  $A/Z$ and redo the calculation. We get a  bad fit (figure(\ref{fig3})).
  There is
 an over all deficit in the Nitrogen population and an excess in the boron population.
 Also we note that within a particular $Z$ there is a deficit in
 the neutron rich isotopes. We try to remedy this situation by increasing the $A/Z$ ratio.
 The best fit at this temperature is obtained at an $A/Z=2.3$ (figure(\ref{fig4})),
  but we still obtain an overal deficit in the Nitrogen population; the Carbon 
  fit is good, but there still remains an excess in the boron isotopes especially in the
  neutron rich isotopes.

 On inspection of the fits (figures (\ref{fig1}) to (\ref{fig4})), we note that the 
 best fit is obtained at figure(\ref{fig2}). In  this fit $T=5.0MeV$, $A/Z = 2.2$,
 $V_{fr}/V_{0}=5.5$, and $\log( {\cal H}) = 6.34$. In this figure we note that,
 for the Boron populations we get an excellent agreement with the data. In
 this case there seems to be little change after single decay.
  For the Carbon isotopes the agreement is good. For Nitrogen, we have a good fit
 except for the case of $^{13}N$. 

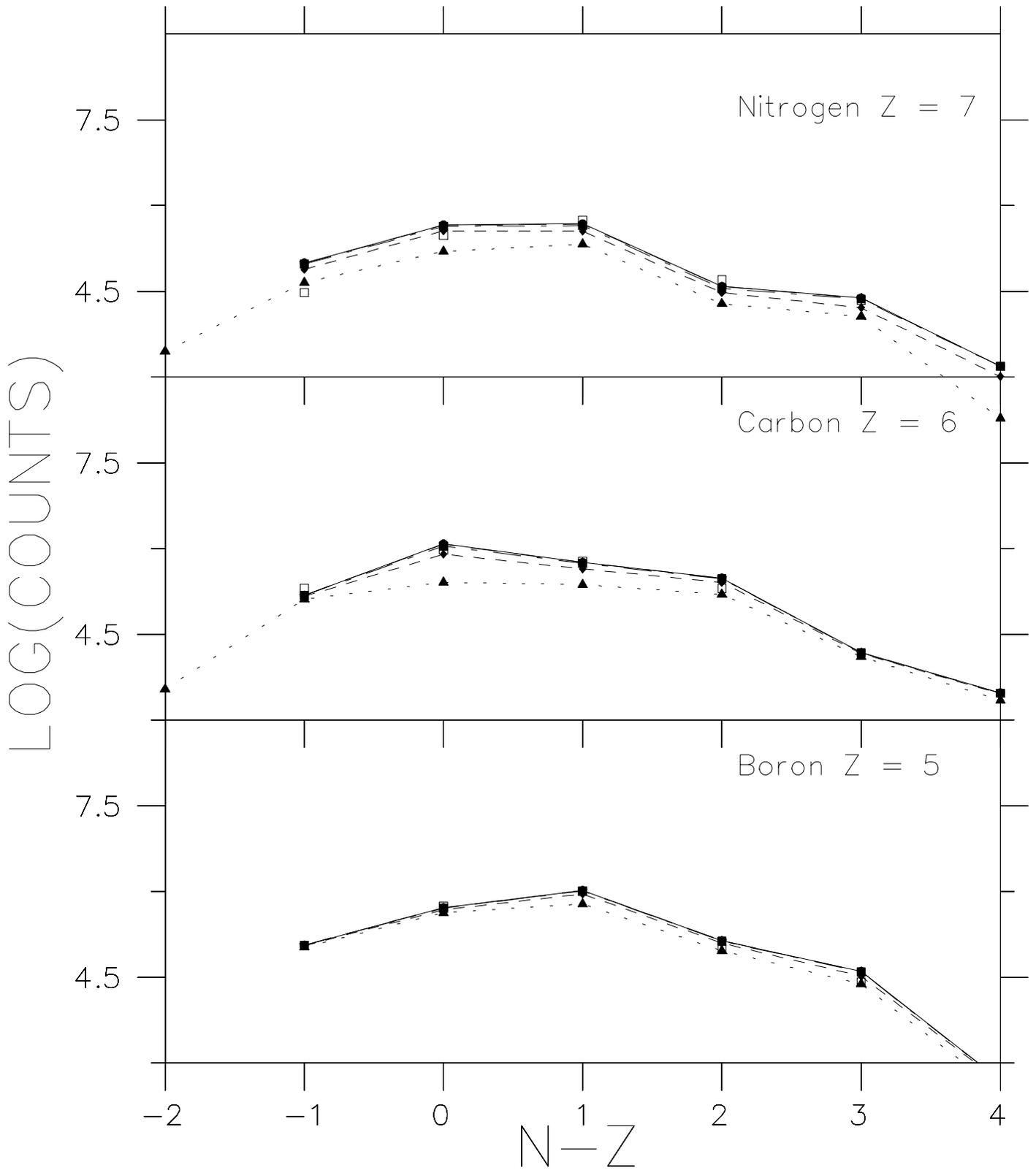
\begin{figure} \caption{ Same as fig.2 but with $Z_{T}=63$, $A_{T}=140$, $T=5.0MeV$,
 $V_{fr}/V_{0}= 5.5$ and $log{\cal H } = 6.30 $. } \label{fig9} \end{figure}

Another property of the fits noticed is that they do not seem to depend on $A$ and $Z$
independently but rather on the ratio $A/Z$. As a demonstration of this, we plot in
figure(\ref{fig9}) a fit for $A=140$ and $Z=63$ ( i.e., $A/Z=2.22$ ). We note that we are 
able
to obtain a fit very similar to fig.\ref{fig2}, with the same temperature and
$V_{fr}/V_{0}$ as in fig.\ref{fig2}, but with a slightly lower ${\cal H}$. This is very 
much expected, as in this case each source has a larger number of nucleons than before.

\section { DISCUSSONS AND CONCLUSION . } 
\vspace{1cm}

In this note we have presented a secondary decay formalism and performed 
calculations to fit the populations of various isotopes measured in \cite{xu89}. 
We obtain very good fits (fig.\ref{fig2}) with
experiment for the Boron and Carbon isotopes. In the Nitrogen isotopes,
 we obtain a good fit except for the case of $^{13}N$. No particular reason could be 
 found for this, but 
let us go over several approximations ( introduced to keep the calculation at a 
resonably simple level )  which may have contributed. 

Actual energy levels from data tables were used only upto $A=20$ (equation (\ref{qreal})) 
for the primary
populations. For higher masses, the emperical mass formula (equation (\ref{qint})) was
 used. The secondary decay is very approximate, instead of calculating decay 
level to level, we have blurred out such details by using a smoothed level density.

For the capture cros-section (equation(\ref{cros})), 
we have used a simple semiclassical formula, assuming that all nuclei are spherically
symmetric which is definitely not true. A more precise calculation involving level to
level decay would use a more accurate expression for the cros-sections 
e.g., the  Hauser$-$Feshbach formalism \cite{hau52} \cite{xi99}. 

Still another problem lies in the assumption made in calculating the effects of higher
order decay, that the interim populations can be taken to be canonically distributed. 
This is true only in first order decay, thus
making the higher order contributions subject to some error. 

There is also an experimental problem according to the authors of \cite{xu89}, the
angular distributions were forward peaked, indicating significant emission prior to
attainment of thermodynamic equilibrium. Such an emission could affect the populations
of the various isotopes.

No doubt, incorporating changes to correct the above mentioned problems will improove 
the accuracy of the calculation. However, such changes may make the
expressions analytically intractible and one would have to resort to numerical means.
This may slow down the calculation considerably.  The calculations presented 
in this note take minimal computer time.  Inspite of the shortcomings of the 
calculation presented above, this still remains a good test of the two component
statistical model, and shows that such a model can definintely be used to explain
certain experimental data quite accurately.

\section {ACKNOWLEDGMENT}

\vspace{1cm}

The authors wish to thank C. Gale, W. G. Lynch, M. B. Tsang, S. Pratt and 
N. de Takacsy for helpful discussions. This work was supported in part by the Natural
Sciences and Engineering Research Council of Canada and by { \it le fonds pour la
Formation de Chercheurs et l'aide \'{a} la Recherche du Qu\'{e}bec. }

\vspace{3cm}

\section {APPENDIX. }

\vspace{1cm}

 \textbf{ (i) Derivation of equation \ref{big}. }

The full decay rate from a particular energy level $E$ of a nucleus 
$A$ (or $(i,j)$) which is decaying by emitting a particle $a$ (or $(x,y)$), 
is given as \ref{befb}, 

\begin{equation}
\Gamma_{a} = \int_{V_{a}}^{E-B_{a}} W(E,\varepsilon) d\varepsilon
\end{equation}

\noindent where $W(E,\varepsilon)$ is the Weisskopf decay probability per unit time 
given by equation (\ref{wisko}). On writing down the full expression for $W$ we get 

\begin{equation}
\Gamma_{a} = \int_{V_{a}}^{E-B_{a}} \frac {\gamma_{a} ^{\prime} \rho_{B}^{0} }{
\rho_{A}(E) \rho_{A}^{0}}
(\varepsilon-V_{a}) \exp \left[ \frac{\pi}{\sqrt{\epsilon_{F}}} \left\{ 
\sqrt{(i+j-x-y) (E-B_{a}-\varepsilon) } \right\} \right] d\varepsilon  
\end{equation}

\noindent Now we substitute $z = (\sqrt { E-B_{a}-\varepsilon })$ and integrate over $z$ 
and let $ C = \left( \frac{ \pi}{ \sqrt{ \epsilon_{F} }} \sqrt{i+j-x-y} \right) $, then 

\begin{equation}
\Gamma_{a} (E)= \int_{0}^{\sqrt{E-B_{a}-V_{a}}} \frac{\gamma_{a} ^{\prime} \rho_{B}^{0}}
{\rho_{A}(E) \rho_{A}^{0}} 
2z(E-B_{a}-V_{a}-z^{2}) \exp (C z) dz 
\end{equation}
 
\noindent on carrying out this simple integration we get,

\begin{eqnarray} 
\Gamma_{a} (E) &=& \frac { 2 \gamma_{a}\prime \rho_{B}^{0}}{ \rho_{A}(E) \rho_{A}^{0}} 
\left[ (E-B_{a}-V_{a}) \left(
\frac { e^{C {\cal B} } }{C} ( {\cal B} - 1/C) - \frac { e ^{C {\cal A} } }{C} 
( {\cal A} - 1/C) \right) \right. \nonumber \\
 & & \left. \mbox{} - \frac { e ^{C {\cal B} }}{C} \left( {\cal B} ^{3} - \frac 
 {3 {\cal B} ^{2} }{C} + 
 \frac { 6 {\cal B} }{C^{2}} - \frac {6}{C^{3}} \right)\right. \nonumber \\
 & & \left. \mbox{} + \frac {e ^{C { \cal A} }}{C} \left( { \cal A} ^{3}-\frac {3
 {\cal A} ^{2}}{C}+\frac {6 { \cal A} }{C^{2}}-
\frac {6}{C^{3}} \right) \right]  \label{bigagn}
\end{eqnarray}

\noindent with ${\cal B} = \sqrt{E-B_{a}-V_{a}}$ , ${\cal A} = 0$.  The
stable decay rate, i.e., the decay rate from an energy level $E$ of the nucleus
$A$ to any of the allowed stable levels of $B$ is given simply
from the above expression by replacing the upper limit to ${\cal B} = \sqrt{E_{B}^{s}}$
where $E_{B}^{s}$ is the stable threshold of the residue nucleus $B$ .
However in the event that $E-B_{a}-V_{a} \leq E_{B}^{s}$ then the above mentioned 
replacement should not be made. In this case the total decay rate is the same as the
 stable decay rate.

\textbf {(ii) Derivation of the generic expression for 
$_{L_{k}} \langle \Gamma_{a} \rangle _{L_{k+1}} $ } 

From equation(\ref{gamma_lk}) we obtain the definition of $ _{L_{k}} \langle \Gamma_{a} 
\rangle _{L_{k+1}}$ as 

\begin{eqnarray}
_{L_{k}} \langle \Gamma_{a} \rangle _{L_{k+1}} =\int_{L_{k}}^{L_{k+1}} \Gamma _{a} (E)
C_{i,j} \rho (E) e^{-\beta E } dE 
\end{eqnarray}

\noindent now taking the expression of $\Gamma_{a}(E)$ from equation(\ref{bigagn}) and
substituting  $z = {\cal B} = (\sqrt{E-B_{a}-V_{a}})$ we get 

\begin{eqnarray}
_{L_{k}} \langle \Gamma_{a} \rangle _{L_{k+1}} &=&
\int_{{\tiny \sqrt{L_{k}-B_{a}-V_{a}}}}^{{\tiny \sqrt{L_{k+1}-B_{a}-V_{a}}}} 
4\gamma_{a}^{\prime} C_{i,j} \frac{\rho_{B}^{0}}{\rho_{A}^{0}}
\Bigg[ \left\{ \frac{2z^{3}}{C^{2}} - \frac{6z^{2}}{C^{3}} + 
\frac{6z}{C^{4}} \right\} e^{Cz} \nonumber \\ 
  & & \mbox{} + \frac{z^{3}}{C^{2}} - \frac{6z}{C^{4}} \Bigg]
e^{-\beta(z^{2}+B_{a}+V_{a})} dz \label{generic}
\end{eqnarray}

\noindent now we may separate the integration into three parts

\begin{equation}
_{L_{k}} \langle \Gamma_{a} \rangle _{L_{k+1}} = \left(  I_{L_{k},L_{k+1}}^{1} +                  
 I_{L_{k},L_{k+1}}^{2} + I_{L_{k},L_{k+1}}^{3} \right) \frac{\rho_{B}^{0}}{\rho_{A}^{0}}
\end{equation}

\noindent where, 

\begin{equation}
I_{L_{k},L_{k+1}}^{1} = \int_{\sqrt{L_{k}-B_{a}-V_{a}}}^{\sqrt{L_{k+1}-B_{a}-V_{a}}}-
C_{i,j} 24 \gamma_{a}^{\prime}  \frac{z}{C^{4}} e^{-\beta(z^{2}+B_{a}+V_{a})}dz
\end{equation}

\begin{equation}
I_{L_{k},L_{k+1}}^{2} = \int_{\sqrt{L_{k}-B_{a}-V_{a}}}^{\sqrt{L_{k+1}-B_{a}-V_{a}}}
C_{i,j} 4 \gamma_{a}^{\prime}  \frac{z^{3}}{C^{4}} e^{-\beta(z^{2}+B_{a}+V_{a})}dz
\end{equation}

\begin{equation}
I_{L_{k},L_{k+1}}^{3} = \int_{\sqrt{L_{k}-B_{a}-V_{a}}}^{\sqrt{L_{k+1}-B_{a}-V_{a}}}
2 \gamma_{a}^{\prime}  C_{i,j} \left\{
\frac{4z^{3}}{C^{2}} - \frac{12z^{2}}{C^{3}} + \frac{12z}{C^{4}}
\right\}e^{Cz-\beta(z^{2}+B_{a}+V_{a})} dz
\end{equation}

\noindent The three integrals can be done simply to give 

\begin{equation}
I_{L_{k},L_{k+1}}^{1} = \frac { -12 \gamma_{a}^{\prime}  C_{i,j} }{ C^{4} \beta } \left[ e^{ - 
\beta L_{k} } - e^{ - \beta L_{k+1} } \right]
\end{equation}

\begin{equation}
I_{L_{k},L_{k+1}}^{2} = \frac { 2 \gamma_{a}^{\prime}  C_{i,j} }{ C^{2} \beta } \left[ L_{k}e^{ - 
\beta L_{k} } - L_{k+1}e^{ - \beta L_{k+1}} + \left\{ 1/ \beta - B_{a} - V_{a} \right\} 
\left( e^{ - \beta L_{k} } - e^{ - \beta L_{k+1}} \right) \right]
\end{equation}

\begin{eqnarray}
I_{L_{k},L_{k+1}}^{3} &=& \frac {4 \gamma_{a}^{\prime}  C_{i,j} }{ C^{2} \beta } e^{- \beta ( B_{a}
 + V_{a} )} e^{ \frac {C^{2}}{4 \beta ^{2} } } \Bigg[ M_{k}^{2} e^{- \beta M_{k}^{2} } -
 M_{k+1}^{2} e^{ -\beta M_{k+1}^{2} }   \nonumber \\
 & & \mbox{} + e^{- \beta M_{k}^{2} } - e^{ -\beta M_{k+1}^{2} } + \left( \frac {3 C^{2} }{2 
 \beta }
- \frac {3}{C} \right) \left\{ M_{k} e^{- \beta M_{k}^{2} } - M_{k+1} e^{ -\beta M_{k+1}^{2}
} \right\}  \mbox{} \nonumber \\
& & \mbox{} + \left( \frac { 3C^{2}}{ 4 \beta^{2} } - \frac {3}{ \beta } + \frac {3}{ C^{2}}
\right) \left\{ e^{- \beta M_{k}^{2} } - e^{ -\beta M_{k+1}^{2} } \right\} \nonumber \\ 
 & & \mbox{}+ \frac { C^{3}
\sqrt{ \pi } }{ 8 \beta ^{2} } \left\{ erf( \sqrt{ \beta } M_{k+1}) - 
erf( \sqrt{ \beta } M_{k}) \right\} \Bigg ] 
\end{eqnarray}

where $M_{k} = \sqrt{ L_{k} - B_{a} - V_{a} } - C/(2 \beta )$ and $M_{k+1} = \sqrt{ 
L_{k+1} - B_{a} - V_{a} } - C/(2 \beta )$ and $C$ is the same as in equation 
\ref{bigagn}. 

The calculation of the stable decay rate is a bit more involved in the limits of
integration and three cases emerge. If  $L_{k} - B_{a} - V_{a} < E_{B}^{s} $, and 
$L_{k+1} - B_{a} - V_{a} \leq E_{B}^{s} $, then 

\begin{equation}
_{L_{k}} \langle \Gamma_{a}^{s} \rangle _{L_{k+1}} = _{L_{k}} \langle \Gamma_{a}
\rangle _{L_{k+1}}
\end{equation}

\noindent if  $L_{k} - B_{a} - V_{a} < E_{B}^{s} $, but $L_{k+1} - B_{a} - V_{a} >
E_{B}^{s} $, then the calculation of $_{L_{k}} \langle \Gamma_{a}^{s} \rangle _{L_{k+1}}$
has to be done in two parts 

\begin{equation}
_{L_{k}} \langle \Gamma_{a}^{s} \rangle _{L_{k+1}} = \left( I^{s}_{1} + I^{s}_{2} \right)
\frac{\rho_{B}^{0}}{\rho_{A}^{0}}
\end{equation}

\noindent where

\begin{equation}
I^{s}_{1} = \int_{L_{k}}^{E_{B}^{s}+B_{a}+V_{a}} \Gamma_{a}(E) C_{i,j} \rho(E)
e^{-\beta E} dE
\end{equation}
 
\noindent the expression for this is the same as equation(\ref{generic}) with the
appropriate change of limits. 

\begin{equation}
I^{s}_{2} =\int_{E_{B}^{s}+B_{a}+V_{a}}^{L_{k+1}} \Gamma_{a}^{s}(E) C_{i,j} \rho(E)
e^{-\beta E} dE
\end{equation}

\noindent if however, $L_{k} - B_{a} - V_{a} \geq E_{B}^{s} $, and $L_{k+1} - B_{a} - V_{a}
> E_{B}^{s} $, then 

\begin{equation}
_{L_{k}} \langle \Gamma_{a}^{s} \rangle _{L_{k+1}} = \int_{L_{k}}^{L_{k+1}} \Gamma_{a}^{s}
(E) C_{i,j} \rho(E) e^{-\beta E} dE
\end{equation}

\noindent the above two integrals are rather trivial and thus detailed expressions are
not presented.

\end{document}